\documentstyle[twoside]{article}

\catcode`\@=11
\long\def\@makefntext#1{
\protect\noindent \hbox to 3.2pt {\hskip-.9pt  
$^{{\eightrm\@thefnmark}}$\hfil}#1\hfill}		

\def\@makefnmark{\hbox to 0pt{$^{\@thefnmark}$\hss}}	
	
\def\ps@myheadings{\let\@mkboth\@gobbletwo
\def\@oddhead{\hbox{}
\rightmark\hfil\eightrm\thepage}   
\def\@oddfoot{}\def\@evenhead{\eightrm\thepage\hfil
\leftmark\hbox{}}\def\@evenfoot{}
\def\sectionmark##1{}\def\subsectionmark##1{}}

\oddsidemargin=\evensidemargin
\newcounter{sectionc}\newcounter{subsectionc}\newcounter{subsubsectionc}
\renewcommand{\section}[1] {\vspace{12pt}\addtocounter{sectionc}{1} 
\setcounter{subsectionc}{0}\setcounter{subsubsectionc}{0}\noindent 
	{\tenbf\thesectionc. #1}\par\vspace{5pt}}
\renewcommand{\subsection}[1] {\vspace{12pt}\addtocounter{subsectionc}{1} 
	\setcounter{subsubsectionc}{0}\noindent 
	{\bf\thesectionc.\thesubsectionc. {\kern1pt \bfit #1}}\par\vspace{5pt}}
\renewcommand{\subsubsection}[1] {\vspace{12pt}\addtocounter{subsubsectionc}{1}
	\noindent{\tenrm\thesectionc.\thesubsectionc.\thesubsubsectionc.
	{\kern1pt \tenit #1}}\par\vspace{5pt}}
\newcommand{\nonumsection}[1] {\vspace{12pt}\noindent{\tenbf #1}
	\par\vspace{5pt}}

\newcounter{appendixc}
\newcounter{subappendixc}[appendixc]
\newcounter{subsubappendixc}[subappendixc]
\renewcommand{\thesubappendixc}{\Alph{appendixc}.\arabic{subappendixc}}
\renewcommand{\thesubsubappendixc}
	{\Alph{appendixc}.\arabic{subappendixc}.\arabic{subsubappendixc}}

\renewcommand{\appendix}[1] {\vspace{12pt}
        \refstepcounter{appendixc}
        \setcounter{figure}{0}
        \setcounter{table}{0}
        \setcounter{lemma}{0}
        \setcounter{theorem}{0}
        \setcounter{corollary}{0}
        \setcounter{definition}{0}
        \setcounter{equation}{0}
        \renewcommand{\thefigure}{\Alph{appendixc}.\arabic{figure}}
        \renewcommand{\thetable}{\Alph{appendixc}.\arabic{table}}
        \renewcommand{\theappendixc}{\Alph{appendixc}}
        \renewcommand{\thelemma}{\Alph{appendixc}.\arabic{lemma}}
        \renewcommand{\thetheorem}{\Alph{appendixc}.\arabic{theorem}}
        \renewcommand{\thedefinition}{\Alph{appendixc}.\arabic{definition}}
        \renewcommand{\thecorollary}{\Alph{appendixc}.\arabic{corollary}}
        \renewcommand{\theequation}{\Alph{appendixc}.\arabic{equation}}
        \noindent{\tenbf Appendix \theappendixc #1}\par\vspace{5pt}}
\newcommand{\subappendix}[1] {\vspace{12pt}
        \refstepcounter{subappendixc}
        \noindent{\bf Appendix \thesubappendixc. {\kern1pt \bfit #1}}
	\par\vspace{5pt}}
\newcommand{\subsubappendix}[1] {\vspace{12pt}
        \refstepcounter{subsubappendixc}
        \noindent{\rm Appendix \thesubsubappendixc. {\kern1pt \tenit #1}}
	\par\vspace{5pt}}

\newcommand{\textlineskip}{\baselineskip=13pt}
\newcommand{\smalllineskip}{\baselineskip=10pt}

\def\eightcirc{
\begin{picture}(0,0)
\put(4.4,1.8){\circle{6.5}}
\end{picture}}
\def\eightcopyright{\eightcirc\kern2.7pt\hbox{\eightrm c}} 

\newcommand{\copyrightheading}[1]
	{\vspace*{-2.5cm}\smalllineskip{\flushleft
	{\footnotesize Modern Physics Letters A, #1} \hfill{nlin.SI/0103055}\\
	{\footnotesize $\eightcopyright$\, World Scientific Publishing
	 Company}\\
	 }}

\newcommand{\publisher}[2]{{\begin{center}\footnotesize\smalllineskip 
	Received #1\\
	Revised #2
	\end{center}
	}}

\def\abstracts#1#2#3{{
	\centering{\begin{minipage}{4.5in}\footnotesize\baselineskip=10pt
	\parindent=0pt #1\par 
	\parindent=15pt #2\par
	\parindent=15pt #3
	\end{minipage}}\par}} 

\renewenvironment{thebibliography}[1]
	{\frenchspacing
	 \ninerm\baselineskip=11pt
	 \begin{list}{\arabic{enumi}.}
        {\usecounter{enumi}\setlength{\parsep}{0pt}     
	 \setlength{\leftmargin 12.7pt}{\rightmargin 0pt} 
         \setlength{\itemsep}{0pt} \settowidth
	{\labelwidth}{#1.}\sloppy}}{\end{list}}

\newcounter{itemlistc}
\newcounter{romanlistc}
\newcounter{alphlistc}
\newcounter{arabiclistc}

\newcommand{\fcaption}[1]{
        \refstepcounter{figure}
        \setbox\@tempboxa = \hbox{\footnotesize Fig.~\thefigure. #1}
        \ifdim \wd\@tempboxa > 5in
           {\begin{center}
        \parbox{5in}{\footnotesize\smalllineskip Fig.~\thefigure. #1}
            \end{center}}
        \else
             {\begin{center}
             {\footnotesize Fig.~\thefigure. #1}
              \end{center}}
        \fi}

\newcommand{\tcaption}[1]{
        \refstepcounter{table}
        \setbox\@tempboxa = \hbox{\footnotesize Table~\thetable. #1}
        \ifdim \wd\@tempboxa > 5in
           {\begin{center}
        \parbox{5in}{\footnotesize\smalllineskip Table~\thetable. #1}
            \end{center}}
        \else
             {\begin{center}
             {\footnotesize Table~\thetable. #1}
              \end{center}}
        \fi}

\def\@citex[#1]#2{\if@filesw\immediate\write\@auxout
	{\string\citation{#2}}\fi
\def\@citea{}\@cite{\@for\@citeb:=#2\do
	{\@citea\def\@citea{,}\@ifundefined
	{b@\@citeb}{{\bf ?}\@warning
	{Citation `\@citeb' on page \thepage \space undefined}}
	{\csname b@\@citeb\endcsname}}}{#1}}

\newif\if@cghi
\def\cite{\@cghitrue\@ifnextchar [{\@tempswatrue
	\@citex}{\@tempswafalse\@citex[]}}
\def\citelow{\@cghifalse\@ifnextchar [{\@tempswatrue
	\@citex}{\@tempswafalse\@citex[]}}
\def\@cite#1#2{{$\null^{#1}$\if@tempswa\typeout
	{IJCGA warning: optional citation argument 
	ignored: `#2'} \fi}}

\def\pmb#1{\setbox0=\hbox{#1}
	\kern-.025em\copy0\kern-\wd0
	\kern.05em\copy0\kern-\wd0
	\kern-.025em\raise.0433em\box0}

\def\fnt#1#2{\footnotetext{\kern-.3em
	{$^{\mbox{\scriptsize #1}}$}{#2}}}

\def\fpage#1{\begingroup
\voffset=.3in
\thispagestyle{empty}\begin{table}[b]\centerline{\footnotesize #1}
	\end{table}\endgroup}

\def\runninghead#1#2{\pagestyle{myheadings}
\markboth{{\protect\footnotesize\it{\quad #1}}\hfill}
{\hfill{\protect\footnotesize\it{#2\quad}}}}
\font\tenrm=cmr10
\font\tenit=cmti10 
\font\tenbf=cmbx10
\font\bfit=cmbxti10 at 10pt
\font\ninerm=cmr9

\font\eightrm=cmr8

\def\qed{\hbox{${\vcenter{\vbox{			
   \hrule height 0.4pt\hbox{\vrule width 0.4pt height 6pt
   \kern5pt\vrule width 0.4pt}\hrule height 0.4pt}}}$}}

\def\rf#1{(\ref{eq:#1})}
\def\lab#1{\label{eq:#1}}
\def\nonu{\nonumber}
\def\br{\begin{eqnarray}}
\def\er{\end{eqnarray}}
\def\be{\begin{equation}}
\def\ee{\end{equation}}

\def\foot#1{\footnotemark\footnotetext{#1}}
\def\lb{\lbrack}
\def\rb{\rbrack}

\def\llb{\left\lbrack}
\def\rrb{\right\rbrack}

\def\lcurl{\left\{}
\def\rcurl{\right\}}
\def\({\left(}
\def\){\right)}
\def\bv{\bigm\vert}               

\def\bc{\begin{center}}
\def\ec{\end{center}}

\newcommand\partder[2]{{{\partial {#1}}\over{\partial {#2}}}}

\newcommand\Sbr[2]{\Bigl\lbrack\,{#1}\, ,\,{#2}\,\Bigr\rbrack} 
 
\def\d{\delta}

\def\vareps{\varepsilon}

\def\h{{1\over 2}}

\def\l{\lambda}
\def\L{\Lambda}

\def\o{\over}

\def\vp{\varphi}
\def\P{\Phi}
\def\pa{\partial}

\def\t{\tau}
\def\th{\theta}

\def\wti{\widetilde}

\newcommand\twomat[4]{\left(\begin{array}{cc}  
{#1} & {#2} \\ {#3} & {#4} \end{array} \right)}
\newcommand\threemat[9]{\left(\begin{array}{ccc}  
{#1} & {#2} & {#3}\\ {#4} & {#5} & {#6}\\
{#7} & {#8} & {#9} \end{array} \right)}

\def\cA{{\cal A}}
\def\cB{{\cal B}}

\def\cD{{\cal D}}

\def\cF{{\cal F}}

\def\cL{{\cal L}}
\def\cM{{\cal M}}

\def\cT{{\cal T}}

\def\cW{{\cal W}}

\def\cZ{{\cal Z}}

\font\msb=msbm10 scaled \magstep1

\newcommand{\IZ}{\mbox{\msb Z} }

\newfam\euffam
\font\sixeuf=eufm6
\font\eighteuf=eufm8
\font\twelveeuf=eufm10 scaled\magstep1
 \textfont\euffam=\twelveeuf
 \scriptfont\euffam=\eighteuf
   \scriptscriptfont\euffam=\sixeuf

\def\one{\hbox{{1}\kern-.25em\hbox{l}}}
\def\0#1{\relax\ifmmode\mathaccent"7017{#1}%
        \else\accent23#1\relax\fi}
\def\Wino{{\bf W_{1+\infty}}}           

\newcommand\DB{{Darboux-B\"{a}cklund}~}

\newcommand\st[2]{\stackrel{(#1 )}{#2}}

\newcommand\cSKPe{${\sl SKP}^{N=2}_{(M_B,M_F)}$~}

\newcommand{\ct}[1]{\cite{#1}}
\newcommand{\bi}[1]{\bibitem{#1}}
\newcommand\NPB[3]{{\sl Nucl. Phys.} {\bf B#1}, #3 (#2)}

\newcommand\CMP[3]{{\sl Commun. Math. Phys.} {\bf #1}, #3 (#2)}

\newcommand\PLA[3]{{\sl Phys. Lett.} {\bf #1A}, #3 (#2)}
\newcommand\PLB[3]{{\sl Phys. Lett.} {\bf #1B}, #3 (#2)}

\newcommand\JMP[3]{{\sl J. Math. Phys.} {\bf #1}, #3 (#2)}

\newcommand\LMP[3]{{\sl Letters in Math. Phys.} {\bf #1}, #3 (#2)}
\newcommand\IJMPA[3]{{\sl Int. J. Mod. Phys.} {\bf A#1}, #3 (#2)}

\newcommand\JPA[3]{{\sl J. Physics} {\bf A#1}, #3 (#2)}

\newcommand\MPLA[3]{{\sl Mod. Phys. Lett.} {\bf A#1}, #3 (#2)}

\begin{document}
 
\vskip .2in

\runninghead{N=2 Supersymmetric Integrable Hierarchies
$\ldots$}{N=2 Supersymmetric Integrable Hierarchies
$\ldots$} 

\normalsize\textlineskip
\thispagestyle{empty}
\setcounter{page}{1}

\copyrightheading{}			

\vspace*{0.88truein}

\fpage{1}
 
\centerline{\bf N=2 Supersymmetric Integrable Hierarchies:}
\centerline{\bf Additional Symmetries and \DB Solutions}

\vspace*{0.37truein}
\centerline{\footnotesize E. Nissimov${}^\ast$ and S. Pacheva\footnote{E-mail: 
nissimov@inrne.bas.bg , svetlana@inrne.bas.bg}}
\vspace*{0.015truein}
\centerline{\footnotesize\it Institute for Nuclear Research and Nuclear Energy, 
Bulgarian Academy of Sciences}
\baselineskip=10pt
\centerline{\footnotesize\it Boul. Tsarigradsko Chausee 72, BG-1784 ~Sofia,
Bulgaria}
\vspace*{0.225truein}
\publisher{(received date)}{(revised date)}

\vspace*{0.21truein} 
\abstracts{We study additional non-isospectral symmetries of constrained
(reduced) $N\! =\! 2$ supersymmetric KP hierarchies of integrable
``soliton''-like evolution equations. These symmetries are shown to form an
infinite-dimensional non-Abelian superloop superalgebra. Furthermore we
study the general \DB (DB) transformations (including adjoint-DB and binary
DB) of $N\! =\! 2$ super-KP hierarchies preserving (most of) the additional 
symmetries. Also we derive the explicit form of the general DB ($N\! =\! 2$ 
``super-soliton''-like) solutions in the form of generalized Wronskian-like
super-determinants.}{}{}

\vspace{.1in}

\vspace*{1pt}\textlineskip
\noindent
\section{Introduction}
\vspace*{-0.5pt}
\noindent
The principal importance of supersymmetric integrable systems from the
point of view of theoretical physics is rooted in their intimate relevance
for (multi-)matrix models in non-perturbative superstring theory
\ct{SI-sstring}. Accordingly, from purely mathematical point of view, the
subject of supersymmetrization of Kadomtsev-Petviashvili (KP) integrable 
hierarchy of ``soliton''-like evolution equations and its various reductions
\ct{SKP-1,SKP-2,SKP-Hen} similarly attracted a lot of interest, especially, the 
supersymmetric generalizations of the inverse scattering method, bi-Hamiltonian 
structures, tau-functions and Sato Grassmannian approach, and the
Drinfeld-Sokolov scheme.

An important role in the theory of integrable systems is being played by the
notion of {\em additional (non-isospectral) symmetries} \ct{Fuchs-Orlov-Schulman}
(for detailed reviews of the latter subject, see refs.\ct{addsym-review}).
Additional symmetries, by definition, consist of the set of all flows on the 
space of the Lax operators of the pertinent integrable hierarchy which commute 
with the ordinary isospectral flows, the latter being generated by the complete
set of commuting integrals of motion. One of the most significant
manifestations of additional symmetries is the interpretation of the crucial 
Virasoro (and $\Wino$) constraints on partition functions of (multi-)matrix 
models of string theory as invariance of the $\t$-functions ({\sl i.e.}, 
the string partition functions) under the Borel subalgebra of the Virasoro 
algebra of additional non-isospectral symmetries in the underlying integrable 
hierarchies (similarly for the $\Wino$ constraints). 

Furthermore, the notion of additional symmetries allows to provide an alternative
formulation of matrix (multi-component) KP hierarchies as ordinary scalar KP 
hierarchy supplemented with appropiate sets of mutually commuting additional 
symmetry flows (see refs.\ct{multi-comp-KP,hallifax}). In particular, one 
obtains an alternative formulation of various physically relevant nonlinear
evolution equations in two- and higher-dimensional space-time
({\sl e.g.}, Davey-Stewartson and $N$-wave resonant systems, as well as 
Wess-Zumino-Novikov-Witten models of group-coset-valued fields)
as additional-symmetry flows on ordinary (reduced) KP hierarchies
(see refs.\ct{multi-comp-KP,hallifax,virflow,gauge-wz}).

Finally, additional non-isospectral symmetries are expected to play
an important role in the quantization of integrable systems (since the
quantum state space must carry representations of the underlying additional
symmetry algebra).

The first main topic of the present Letter is the study of additional 
non-isospectral symmetries of constrained (reduced) $N\! =\! 2$ supersymmetric 
KP integrable hierarchies. This is an extension of our study in
ref.\ct{svirflow} of additional symmetries of $N\! =\! 1$ super-KP hierarchies.
These symmetries are shown to form an infinite-dimensional non-Abelian superloop
superalgebra. The second main topic is the study of general \DB (DB) 
transformations (including adjoint-DB and binary DB) of $N\! =\! 2$ super-KP 
hierarchies preserving (most of) the additional symmetries. The main result
here is the derivation of the explicit form of the general DB ($N\! =\! 2$ 
``super-soliton''-like) solutions in the form of generalized Wronskian-like
super-determinants. An extended exposition of the present results, including 
detailed proofs and further developments, will appear elsewhere.

\section{Superspace Formulation of General N=2 Super-KP Hierarchy}
%
\noindent
We shall use throughout the $N\! =\! 2$ super-pseudo-differential operator calculus
as in ref.\ct{N2-SKP} with the following notations: 
\be
\pa \equiv \partder{}{x} \quad ,\quad 
\cD_{\pm} = \partder{}{\th_{\pm}} + \th_{\pm} \partder{}{x} \quad; \quad
\cD_{\pm}^2 = \pa \quad ,\quad \lcurl \cD_{+},\,\cD_{-}\rcurl = 0
\lab{N2-notat}
\ee
where $(x,\th_{+},\th_{-})$ denote $N\! =\! 2$ superspace coordinates.
For any $N\! =\! 2$ super-pseudo-differential operator $\cA$ :
\be
\cA = \cA_{+} + \cA_{-} \quad ,\quad
\cA_{\pm} \equiv \sum_{j\geq 0} \( a^{(0)}_{\pm j} + a^{(+)}_{\pm j} \cD_{+} +
a^{(-)}_{\pm j} \cD_{-} + a^{(1)}_{\pm j} \cD_{+}\cD_{-}\)\pa^{\pm j}
\lab{N2-oper}
\ee
the subscripts $(\pm )$ denote its purely differential or purely
pseudo-differential parts, respectively. The rules of conjugation within the
super-pseudo-differential formalism are as follows (cf. first ref.\ct{SKP-Hen}):
$(\cA \cB )^\ast = (-1)^{|A|\, |B|} \cB^\ast \cA^\ast$
for any two elements with Grassmann parities $|A|$ and $|B|$;
$\(\pa^k\)^\ast = (-1)^k \pa^k\, ,\,\(\cD_{\pm}^k\)^\ast = (-1)^{k(k+1)/2} \cD_{\pm}^k$
and $u^\ast = u$ for any coefficient superfield.
Also, in order to avoid confusion we shall also employ the following 
notations: for any super-(pseudo-)\-differential operator $\cA$ and a 
superfield function $f$, the symbol $\, \cA (f)\,$ or $\, (\cA f)\,$
will indicate application (action) of $\cA$ on $f$, whereas the
symbol $\cA f$ without brackets will denote just operator product of $\cA$ with the zero-order
(multiplication) operator $f$.

The general unconstrained $N\! =\! 2$ supersymmetric KP hierarchy is given by a
{\em fermionic} $N\! =\! 2$ super-pseudo-differential Lax operator $\cL$ in
terms of the bosonic  $N\! =\! 2$ superspace dressing operator $\cW$ \ct{N2-SKP} :
\be
\cL = \cW \cD_{-} \cW^{-1} \quad ,\quad
\cW \equiv 1 + \sum_{k=1}^\infty \( w^{(0)}_k + w^{(+)}_k \cD_{+}
+ w^{(-)}_k \cD_{-} + w^{(1)}_k \cD_{+}\cD_{-}\) \pa^{-k}
\lab{N2-superlax}
\ee
obeying bosonic and fermionic evolution (isospectral) Sato
equations\foot{In fact, as shown in \ct{N2-SKP}, there exist two additional 
bosonic and two additional fermionic sets of flows such that the $N\! =\! 2$
super-KP hierarchy exibits $N\! =\! 4$ supersymmetry.} :
\be
\partder{}{t_l} \cL = \Sbr{\(\cL^{2l}\)_{+}}{\cL}   \quad ,\quad
D^{-}_n \cL = - \lcurl \(\cL^{2n-1}\)_{-},\,\cL \rcurl    \quad ,\quad
D^{+}_n \cL = \lcurl \(\L^{2n-1}\)_{+},\,\cL \rcurl
\lab{N2-isospec-eqs}
\ee
where $\L = \cW \cD_{+} \cW^{-1}$. The fermionic isospectral flows $D^{\pm}_n$
in Eqs.\rf{N2-isospec-eqs} possess natural realization in terms of two infinite
sets of fermionic ``evolution'' parameters $\lcurl \rho^{\pm}_n\rcurl_{n=1}^\infty$
and span $N\! =\! 2$ supersymmetry algebra :
\be
D^{\pm}_n = \partder{}{\rho^{\pm}_n} - 
\sum_{k=1}^\infty \rho^{\pm}_k \partder{}{t_{n+k-1}} \quad ,\quad
\lcurl D^{\pm}_n,\, D^{\pm}_m \rcurl = - 2 \partder{}{t_{n+m-1}} \quad ,\quad
\lcurl D^{\pm}_n,\, D^{\mp}_m \rcurl = 0
\lab{N2-isospec-alg}
\ee
the rest of flow commutators being zero.

The super-Baker-Akhiezer (super-BA) and the adjoint super-BA wave functions
are defined as ~
$\psi_{BA} = \cW \bigl(\st{0}{\psi}\!\!{}_{BA}\bigr)$ and
$\psi^{\ast}_{BA} = {\cW^\ast}^{-1} \bigl(\st{0}{\psi^\ast}\!\!{}_{BA}\bigr)$
in terms of the ``free'' super-BA functions (with $\eta_{\pm}$ being fermionic
``spectral'' parameters):
\be
\st{0}{\psi}\!\!{}_{BA} (t,\th_{\pm},\rho^{\pm} ;\l ,\eta_{\pm}) 
= e^{\xi (t,\th_{\pm},\rho^{\pm} ;\l ,\eta_{\pm})} \;\; ,\;\;
\st{0}{\psi^\ast}\!\!{}_{BA} (t,\th_{\pm},\rho^{\pm} ;\l ,\eta_{\pm}) 
= e^{-\xi (t,\th_{\pm},\rho^{\pm} ;\l ,\eta_{\pm})}
\lab{free-super-BA}
\ee
\be
\xi (t,\th_{\pm},\rho^{\pm} ;\l ,\eta_{\pm}) \equiv 
\sum_{l=1}^\infty \l^l t_l + 
\sum_{\pm} \Bigl\{\eta_{\pm}\th_{\pm} + (\eta_{\pm} - \l\th_{\pm} )
\sum_{n=1}^\infty \l^{n-1} \rho^{\pm}_n \Bigr\} 
\lab{xi-def}
\ee
for which it holds:
\be  
\partder{}{t_k} \st{0}{\psi}\!\!{}_{BA} = 
\pa^k \st{0}{\psi}\!\!{}_{BA} \quad , \quad
D^{\pm}_n \st{0}{\psi}\!\!{}_{BA} = \cD_{\pm}^{2n-1} \st{0}{\psi}\!\!{}_{BA} = 
\pa_x^{n-1} \cD_{\pm} \st{0}{\psi}\!\!{}_{BA}
\lab{dn-free-super-BA}
\ee
In Eqs.\rf{free-super-BA}--\rf{xi-def} and in what follows we employ the
short-hand notions $(t)\!\equiv\! \( t_1\equiv x,t_2,\right.$\\
$\left. t_3, \ldots\)$ and 
$(\rho^{\pm})\equiv \(\rho^{\pm}_1,\rho^{\pm}_2,\rho^{\pm}_3, \ldots\)$.
Accordingly, (adjoint) super-BA wave functions satisfy: 
\br
\(\cL^2\)^{(\ast)} \psi^{(\ast)}_{BA} = \l \psi^{(\ast)}_{BA} \quad , \quad
\partder{}{t_l} \psi^{(\ast)}_{BA} = 
\pm \(\cL^{2l}\)^{(\ast)}_{+} (\psi^{(\ast)}_{BA})
\nonu
\er
\be
D^{-}_n \psi^{(\ast)}_{BA} = \pm \(\cL^{2n-1}\)^{(\ast)}_{+} (\psi^{(\ast)}_{BA})
\quad , \quad
D^{+}_n \psi^{(\ast)}_{BA} = \pm \(\L^{2n-1}\)^{(\ast)}_{+} (\psi^{(\ast)}_{BA})
\lab{super-BA-eqs}
\ee

Another important object in the present formalism is the notion of $N\! =\! 2$
super-eigenfunctions (SEF's) obeying the following defining equations: 
\be
\partder{}{t_l} \P = \cL^{2l}_{+} (\P) \quad ,\quad
D^{-}_n \P = \cL^{2n-1}_{+} (\P) \quad , \quad
D^{+}_n \P = \L^{2n-1}_{+} (\P) 
\lab{SEF-eqs}
\ee
\be
\partder{}{t_l} \Psi = - \(\cL^{2l}\)^{\ast}_{+} (\Psi) \quad ,\quad
D^{-}_n \Psi = - \(\cL^{2n-1}\)^{\ast}_{+} (\Psi) \quad , \quad
D^{+}_n \Psi = - \(\L^{2n-1}\)^{\ast}_{+} (\Psi)
\lab{adj-SEF-eqs}
\ee
Any (adjoint) SEF possesses unique $N\! =\! 2$ supersymmetric ``spectral''
representation in terms of the (adjoint) super-BA function
similar to the purely bosonic and $N\! =\! 1$ supersymmetric cases 
\ct{ridge,match}: 
\br
\P (t,\th_{\pm},\rho^{\pm}) = \int d\l\, d\eta_{+}\, d\eta_{-}\,
\vp (\l,\eta_{\pm}) \psi_{BA} (t,\th_{\pm},\rho^{\pm};\l ,\eta_{\pm})
\lab{super-spec} \\    
\Psi (t,\th_{\pm},\rho^{\pm}) = \int d\l\, d\eta_{+}\, d\eta_{-}\,
\vp^\ast (\l,\eta_{\pm}) \psi^\ast_{BA}(t,\th_{\pm},\rho^{\pm};\l ,\eta_{\pm})
\nonu
\er
%

%
\section{Constrained N=2 Super-KP Hierarchies}
\noindent
We are now interested in consistent reductions of the general (unconstrained)
$N\! =\! 2$ super-KP hierarchy \rf{N2-superlax}--\rf{N2-isospec-eqs}.
In ref.\ct{N2-SKP} the authors proposed a novel class of constrained
$N\! =\! 2$ super-KP hierarchies defined by $N\! =\! 2$ super-Lax operators
of the form $\cL = \cD_{-} + \P \cD_{+}^{-1}\Psi$ with $\P,\,\Psi$ being in
general matrix-valued superfunctions. On the other hand, we can use the same
logic of the construction of constrained super-KP hierarchies in ref.\ct{match}
to extend it from $N\! =\! 1$ to $N\! =\! 2$ super-KP case. In this way we
obtain the following general class \cSKPe of constrained (reduced) $N\! =\! 2$
super-KP hierarchies defined by $N\! =\! 2$ super-Lax operators:
\be
\cL \equiv \cL_{(M_B,M_F)} = \cD_{-} + \sum_{i=1}^M \P_i \cD_{+}^{-1} \Psi_i 
\lab{Lax-SKP-N2}
\ee
Here $M\! \equiv M_B + M_F$ with $M_{B,F}$ being the number of
bosonic/fermionic superfunctions $\P_i,\, \Psi_i$. Also, for convenience the 
order of indices $i=1,\ldots ,M$ in \rf{Lax-SKP-N2} is chosen such that the 
(adjoint) SEF's $\lcurl \P_i,\,\Psi_i\rcurl_{i=1}^{M_B}$ are bosonic whereas 
the rest $\lcurl \P_{M_B +i},\,\Psi_{M_B +i}\rcurl_{i=1}^{M_F}$ are fermionic.

For later use let us also write down the explicit 
expression of $\(\cL^K\)_{-}$ for arbitrary integer power $K$ of $\cL$ 
\rf{Lax-SKP-N2} \ct{match,N2-SKP}:
\be
\(\cL^{2k}\)_{-} = \sum_{i=1}^M \sum_{s=0}^{2k-1} (-1)^{s|i|}
\cL^{2k-1-s}(\P_i) \cD_{+}^{-1} \(\cL^s\)^\ast (\Psi_i)
\lab{Lax-SKP-N2-2k}
\ee
\br
\(\cL^{2k+1}\)_{-} = \sum_{i=1}^M \sum_{s=0}^{2k} (-1)^{s|i|}
\cL^{2k-s}(\P_i) \cD_{+}^{-1} \(\cL^s\)^\ast (\Psi_i)
\nonu \\
+ \sum_{i=1}^M \sum_{s=0}^{2k-1} (-1)^{s|i|+s+|i|}
\cL^{2k-1-s}(\P_i) \cD_{+}^{-1}\cD_{-} \(\cL^s\)^\ast (\Psi_i)
\lab{Lax-SKP-N2-2k+1}
\er

As in the case of $N\! =\! 1$ constrained super-KP hierarchies \ct{match},
consistency of the action of isospectral flows with the constrained form
\rf{Lax-SKP-N2} of \cSKPe super-Lax operators requires nontrivial 
{\rm modification} of the fermionic $D^{-}_n$ flows \rf{N2-isospec-eqs}. 
The corresponding Sato evolution equations for \rf{Lax-SKP-N2} become:
\br
\partder{}{t_l} \cL = \Sbr{\(\cL^{2l}\)_{+}}{\cL}   \quad ,\quad
D^{+}_n \cL = \lcurl \(\L^{2n-1}\)_{+},\,\cL \rcurl
\nonu \\
D^{-}_n \cL = - \lcurl \(\cL^{2n-1}\)_{-} - X_{2n-1},\,\cL \rcurl 
\lab{SKP-N2-isospec-eqs}
\er
where (cf. Eq.\rf{Lax-SKP-N2-2k+1}):
\be
\(\cL^{2n-1}\)_{-} - X_{2n-1} \equiv \sum_{i=1}^M \sum_{s=0}^{2n-2}
(-1)^{s(|i|+1)} \cL^{2n-2-s}(\P_i) \cD_{+}^{-1} \(\cL^s\)^\ast (\Psi_i)
\lab{X-def}
\ee
Accordingly, the superfunctions $\P_i$ and $\Psi_i$ entering the
pseudo-differential art of $\cL$ \rf{Lax-SKP-N2} are (adjoint) SEF's 
\rf{SEF-eqs}--\rf{adj-SEF-eqs} obeying {\em modified} defining $D^{-}_n$-flow 
equations:
\be
D^{-}_n \P_i = \Bigl(\cL^{2n-1}_{+} + X_{2n-1}\Bigr) (\P_i) - 2\cL^{2n-1}(\P_i)
\lab{Dn-SEF-eqs}
\ee
\be
D^{-}_n \Psi_i = - \Bigl(\(\cL^{2n-1}\)^\ast_{+} + X^\ast_{2n-1}\Bigr) (\Psi_i)
+ 2\(\cL^{2n-1}\)^\ast (\P_i)
\lab{Dn-adj-SEF-eqs}
\ee

Let us point out that, similarly to the $N\! =\! 1$ supersymmetric case,
constrained $N\! =\! 2$ super-KP hierarchies 
\rf{Lax-SKP-N2},\rf{SKP-N2-isospec-eqs} contain among themselves
$N\! =\! 2$ supersymmetric extensions of various basic bosonic integrable
hierarchies such as (modified) Korteveg-de Vries, nonlinear Schr{\"o}dinger
(AKNS hierachies, in general), Yajima-Oikawa, coupled Boussinesq-type
equations etc..

All solutions of (constrained) $N\! =\! 2$ super-KP hierarchies are
expressed through a single $N\! =\! 2$ {\em super-tau function}
~$\t = \t (t,\th_{\pm},\rho^{\pm})$ which is related to the coefficients of
the pertinent $N\! =\! 2$ super-Lax operator $\cL$ \rf{Lax-SKP-N2} and its
associate $\L$ as follows:
\br
\(\cL^{2k}\)_{(-1)} = \partder{}{t_k}\cD_{+} \ln\t  \quad ,\quad
\(\L^{2n-1}\)_{(-1)} = D^{+}_n \cD_{+} \ln\t
\nonu \\
\(\cL^{2n-1} - X_{2n-1}\)_{(-1)} = D^{-}_n \cD_{+} \ln\t
\lab{N2-supertau}
\er
where the subscript $(-1)$ indicates taking the coefficient in front of
$\cD_{+}^{-1}$ in the expansion of the corresponding super-pseudo-differential
operator. The validity of Eqs.\rf{N2-supertau} follows directly from the
$N\! =\! 2$ superspace Zakharov-Shabat ``zero-curvature'' equations, 
{\sl i.e.}, the compatibility conditions among the isospectral Sato evolution
equations \rf{SKP-N2-isospec-eqs} respecting the (anti-)commutation isospectral
flow algebra \rf{N2-isospec-alg}. 

\section{Superloop Superalgebra Symmetries of Constrained N=2 Super-KP Hierarchies}
\noindent
In the study \ct{svirflow} of additional non-isospectral symmetries of 
$N\! =\! 1$ supersymmetric integrable hierarchies we encountered superloop
superalgebras ${\widehat {GL}} (N_1,N_2)$. The latter are 
infinite-dimensional super-Lie algebras with half-integer loop grading 
$0, \pm \h, \pm 1, \pm {3\o 2},\ldots$ :
\be
{\widehat {GL}} (N_1,N_2) = \oplus_{\ell \in \IZ} {GL}^{({\ell \o 2})}(N_1,N_2)
\lab{loop-superalg}
\ee
whose ${\ell \o 2}$-grade subspaces consist of super-matrices of the following form:
\be
{GL}^{(n)}(N_1,N_2) =\lcurl \twomat{A^{(n)}}{B^{(n)}}{C^{(n)}}{D^{(n)}}
\in Mat(N_1,N_2)  \rcurl
\lab{loop-superalg-even}
\ee
\be
{GL}^{(n-\h)}(N_1,N_2) =
\lcurl \twomat{B^{(n-\h)}}{A^{(n-\h)}}{D^{(n-\h)}}{C^{(n-\h)}} \in 
{\wti {Mat}}(N_1,N_2) \rcurl 
\lab{loop-superalg-odd}
\ee
In Eq.\rf{loop-superalg-even} $Mat(N_1,N_2)$ denotes the space of all
$(N_1,N_2) \times (N_1,N_2)$ supermatrices defined in the standard Berezin
block form ({\sl i.e.}, diagonal blocks are bosonic and off-diagonal blocks
are fermionic), whereas the symbol ${\wti {Mat}}(N_1,N_2)$ in
Eq.\rf{loop-superalg-odd} indicates the space of $(N_1,N_2) \times (N_1,N_2)$
supermatrices given in a nonstandard block format with diagonal blocks being
fermionic while the off-diagonal blocks are bosonic (for a detailed
discussion of non-standard formats of super-Lie algebras, see
ref.\ct{Gieres-etal}). Below we will show that the same type of superloop
superalgebras serve as algebras of additional symmetry flows of $N\! =\! 2$
super-KP hierarchies.

Borrowing from the construction in ref.\ct{svirflow} let us consider the
following infinite sets of bosonic and fermionic super-pseudo-differential
operators:
\be
\cM^{(\ell/2)}_\cA \equiv \sum_{i,j=1}^M \cA^{(\ell/2)}_{ij}
\sum_{s=0}^{\ell -1} (-1)^{s(|j| + \ell)} 
\cL^{\ell -1-s}(\P_j) \cD_{+}^{-1} \(\cL^s\)^\ast (\Psi_i)
\lab{M-A-def}
\ee
\be
\cM^{(\ell/2)}_\cF \equiv \sum_{i,j=1}^M \cF^{(\ell/2)}_{ij}
\sum_{s=0}^{\ell -1} (-1)^{s(|j| + \ell)} 
\cL^{\ell -1-s}(\P_j) \cD_{+}^{-1} \(\cL^s\)^\ast (\Psi_i)
\lab{M-F-def}
\ee
where $\ell =1,2,\ldots$ and $\lcurl \P_i,\,\Psi_i\rcurl$ are the same SEF's
entering the pseudo-differential part of $\cL$ \rf{Lax-SKP-N2}.
The associated flows generated by \rf{M-A-def}--\rf{M-F-def} on the
$N\! =\! 2$ constrained super-KP hierarchies \rf{Lax-SKP-N2} read:
\be
\d^{(\ell/2)}_{\cA} \cL = \Sbr{\cM^{(\ell/2)}_{\cA}}{\cL} \quad ,\quad
\d^{(\ell/2)}_{\cF} \cL = \Bigl\{\cM^{(\ell/2)}_{\cF},\,\cL \Bigr\}
\lab{addsym-Lax-SKP-N2}
\ee
or, equivalently, $\d^{(\ell/2)}_{\cA,\cF} \cW = \cM^{(\ell/2)}_{\cA,\cF} \cW$.
In Eqs.\rf{M-A-def}--\rf{M-F-def} 
$\cA^{(\ell/2)}$ and $\cF^{(\ell/2)}$ are constant supermatrices belonging
to the superloop superalgebra ${\widehat {GL}}(M_B,M_F)$ (cf. definition
\rf{loop-superalg}--\rf{loop-superalg-odd}) of the following types:

(a) For $\ell = 2n$ $\cA^{(n)}$ and $\cF^{(n)}$ are purely 
bosonic and purely fermionic elements, respectively:
\be
\cA^{(n)} = \twomat{A^{(n)}}{0}{0}{D^{(n)}} \quad ,\quad  
\cF^{(n)} = \twomat{0}{B^{(n)}}{C^{(n)}}{0}
\lab{A-F-even}
\ee
Here the block matrices $A^{(n)},\, B^{(n)},\ C^{(n)}$ and $D^{(n)}$ are of
sizes $M_B\times M_B$, $M_B\times M_F$, $M_F\times M_B$ and $M_F\times M_F$,
respectively.

(b) For $\ell = 2n -1$ the supermatrices $\cA^{(n-\h)}$ and $\cF^{(n-\h)}$ are
purely bosonic and purely fermionic elements in the ``twisted'' non-standard 
format:
\be
\cF^{(n-\h)} = \twomat{B^{(n-\h)}}{0}{0}{C^{(n-\h)}} \quad ,\quad  
\cA^{(n-\h)} = \twomat{0}{A^{(n-\h)}}{D^{(n-\h)}}{0}
\lab{A-F-odd}
\ee
In this case the sizes of the block matrices $A^{(n-\h)},\, B^{(n-\h)},\ C^{(n-\h)}$ and 
$D^{(n-\h)}$ are $M_B\times M_F$, $M_B\times M_B$, $M_F\times M_F$ and 
$M_F\times M_B$, respectively.

Note from Eq.\rf{M-A-def} that for $\ell =2n$ :
\be
\cM^{(n)}_{\cA=\one} \equiv 
\sum_{j=1}^M \sum_{s=0}^{2n-1} (-1)^{s|j|} 
\cL^{2n-1-s}(\P_j) \cD_{+}^{-1} \(\cL^s\)^\ast (\Psi_j) = \(\cL^{2n}\)_{-}
\lab{Lax-2n-def}
\ee
whereas for $\ell = 2n-1$ Eq.\rf{M-F-def} implies:
\be
\cM^{(n-\h)}_{\cF=\one} \equiv
\sum_{j=1}^M \sum_{s=0}^{2n -2} (-1)^{s(|j|+1)} 
\cL^{2n-2-s}(\P_j)\cD_{+}^{-1}\(\cL^s\)^\ast (\Psi_j) =
\(\cL^{2n-1}\)_{-} - X_{2n-1} 
\lab{Lax-X}
\ee
with $X_{2n-1}$ the same as in Eq.\rf{X-def}.

Since all superfunctions entering the operators $\cM^{(\ell/2)}_{\cA,\cF}$
are SEF's of the constrained $N\! =\! 2$ super-KP hierarchies \rf{Lax-SKP-N2}, 
it is straightforward to show that the flows $\d^{(\ell/2)}_{\cA,\cF}$
\rf{addsym-Lax-SKP-N2} commute with the bosonic isospectral flows
$\partder{}{t_l}$ \rf{N2-isospec-eqs}, thus, identifying \rf{addsym-Lax-SKP-N2}
as additional non-isospectral symmetries of the latter hierarchies.
Furthermore, $\d^{(\ell/2)}_{\cA,\cF}$-flows
(anti-)commute with the second set of fermionic isospectral flows $D^{+}_n$
\rf{N2-isospec-eqs}, whereas the first set of fermionic isospectral flows
$D^{-}_n$ \rf{N2-isospec-eqs} become nontrivial elements of an
infinite-dimensional superloop superalgebra (cf. \rf{super-KM-alg-flows} and
\rf{susy-isospec-flows} below).

In checking the above properties we make use of the following 
super-pseudo-differential operator identities (cf. refs.\ct{match,svirflow}):
\be
\cZ_{(i,j)}\, \cZ_{(k,l)} = \cZ_{(i,j)} (\P_k) \cD_{+}^{-1} \Psi_l +
(-1)^{|j|(|k|+|l|+1)} \P_i \cD_{+}^{-1} \cZ_{(k,l)}^\ast \(\Psi_j\)
\lab{susy-pseudo-diff-id}
\ee
where $\cZ_{(i,j)} \equiv \P_i \cD_{+}^{-1} \Psi_j$ and similarly for
$\cZ_{(k,l)}$.

Consistency of the flow actions \rf{addsym-Lax-SKP-N2} with the constrained 
form of the $N\! =\! 2$ super-Lax operator \rf{Lax-SKP-N2} implies, using
identities \rf{susy-pseudo-diff-id}, the following
actions of $\d^{\ell/2}_{\cA,\cF}$-flows on the constituent (adjoint) SEF's:
\be
\d^{(\ell/2)}_{\cA} \P_i = \cM^{(\ell/2)}_{\cA} (\P_i) -
\sum_{j=1}^M \cA^{(\ell/2)}_{ij}\cL^{\ell}(\P_j) 
\lab{addsym-A-EF}
\ee
\be
\d^{(\ell/2)}_{\cA} \Psi_i = - \(\cM^{(\ell/2)}_{\cA}\)^\ast (\Psi_i) +
\sum_{j=1}^M (-1)^{\ell |j|} \cA^{(\ell/2)}_{ji}\(\cL^{\ell}\)^\ast (\Psi_j)
\lab{addsym-A-adj-EF}
\ee
\be
\d^{(\ell/2)}_{\cF} \P_i = \cM^{(\ell/2)}_{\cF} (\P_i) +
\sum_{j=1}^M \cF^{(\ell/2)}_{ij}\cL^{\ell}(\P_j) 
\lab{addsym-F-EF}
\ee
\be
\d^{(\ell/2)}_{\cF} \Psi_i = - \(\cM^{(\ell/2)}_{\cF}\)^\ast (\Psi_i) -
\sum_{j=1}^M (-1)^{(\ell +1)|j|} \cF^{(\ell/2)}_{ji}
\(\cL^{\ell}\)^\ast (\Psi_j)
\lab{addsym-F-adj-EF}
\ee

Furthermore, employing again identities \rf{susy-pseudo-diff-id}, we obtain:
\be
\d_{\cA_1}^{(\ell/2)} \cM^{(m/2)}_{\cA_2} - 
\d_{\cA_2}^{(m/2)} \cM^{(\ell/2)}_{\cA_1}
- \Sbr{\cM^{(\ell/2)}_{\cA_1}}{\cM^{(m/2)}_{\cA_2}} = 
\cM^{((\ell +m)/2)}_{\lb \cA_1,\cA_2 \rb}
\lab{comm-A1-A2}
\ee
\be
\d_{\cA}^{(\ell/2)} \cM^{(m/2)}_{\cF} - 
\d_{\cF}^{(m/2)} \cM^{(\ell/2)}_{\cA}
- \Sbr{\cM^{(\ell/2)}_{\cA}}{\cM^{(m/2)}_{\cF}} = 
\left\{ \begin{array}{lr}
\cM^{((\ell +m)/2)}_{\lb \cA ,\cF\rb} & {\rm for} \; \ell=even \\
- \cM^{((\ell +m)/2)}_{\{\cA,\cF\}} & {\rm for} \; \ell=odd
\end{array} \right.
\lab{comm-A-F}
\ee
\br
\d_{\cF_1}^{(\ell/2)} \cM^{(m/2)}_{\cF_2} + 
\d_{\cF_2}^{(m/2)} \cM^{(\ell/2)}_{\cF_1}
- \Bigl\{ \cM^{(\ell/2)}_{\cF_1},\,\cM^{(m/2)}_{\cF_2}\Bigr\} = 
\nonu \\
= \left\{ \begin{array}{lr}
\pm \cM^{((\ell +m)/2)}_{\{\cF_1,\cF_2\}} & {\rm for} \;\; 
(\ell,m)=(odd,odd)/(even,even) \\
\pm \cM^{((\ell +m)/2)}_{\lb \cF_1,\cF_2\rb} & {\rm for} \;\; 
(\ell,m)=(odd,even)/(even,odd) \\
\end{array} \right.
\lab{comm-F1-F2}
\er
The super-pseudo-differential operator relations \rf{comm-A1-A2}--\rf{comm-F1-F2},
which constitute the compatibility conditions for the additional symmetry
flows \rf{addsym-Lax-SKP-N2}, imply that the latter obey the following
infinite-dimensional algebra (here $\ell,m \geq 1$) :
\br
\Sbr{\d^{(\ell/2)}_{\cA_1}}{\d^{(m/2)}_{\cA_2}} = 
\d^{((\ell +m)/2)}_{\lb \cA_1,\, \cA_2\rb}
\quad ,\quad 
\Sbr{\d^{(\ell/2)}_{\cA}}{\d^{(m/2)}_{\cF}} =\d^{((\ell +m)/2)}_{\lb \cA,\,\cF\rb}
\;\; {\rm for}\;\; \ell ={\rm even}  \nonu \\
\Sbr{\d^{(\ell/2)}_\cA}{\d^{(m/2)}_\cF} = -\d^{((\ell +m)/2)}_{\{\cA,\,\cF\}}
\;\;\; {\rm for}\;\; \ell ={\rm odd}  \phantom{aaaaaaaaaa}
\nonu
\er
\vspace{-.5cm}
\br
\Bigl\{ \d^{(\ell/2)}_{\cF_1},\, \d^{(m/2)}_{\cF_2} \Bigr\} = 
\pm \d^{((\ell +m)/2)}_{\{ \cF_1,\cF_2\}}
\;\;\; {\rm for}\;\; (\ell,m)=(odd,odd)/(even,even)
\nonu \\
\Bigl\{ \d^{(\ell/2)}_{\cF_1},\, \d^{(m/2)}_{\cF_2} \Bigr\} = 
\pm \d^{((\ell +m)/2)}_{\lb\cF_1,\cF_2\rb}
\;\;\; {\rm for}\;\; (\ell,m)=(odd,even)/(even,odd)
\lab{super-KM-alg-flows}
\er
The algebra \rf{super-KM-alg-flows} is isomorphic to the positive-grade 
subalgebra of the superloop superalgebra ${\widehat {GL}}(M_B,M_F)$.
In particular, from \rf{Lax-2n-def}--\rf{Lax-X} we find that:
\be
\d^{(n)}_{\cA=\one} = - \partder{}{t_n} \quad ,\quad
\d^{(n-\h)}_{\cF=\one} = - D^{-}_n
\lab{susy-isospec-flows}
\ee
are, upto an overall minus sign, isospectral flows of the corresponding
\cSKPe \\ 
hierarchy \rf{Lax-SKP-N2}.

\section{\DB (N=2 ``Super-Soliton'') Solutions}
\noindent
In refs.\ct{gauge-wz,svirflow} we have discussed in detail \DB (DB)
transformations and DB solutions ({\sl i.e.}, soliton-like solutions) of bosonic
and $N\! =\! 1$ supersymmetric constrained KP hierarchies preserving the 
additional non-isospectral symmetries of the latter. Employing the same approach
we are lead to consider the following DB and ajoint-DB transformations of 
$N\! =\! 2$ constrained super-KP hierarchies \cSKPe \rf{Lax-SKP-N2} :
\br
{\wti \cL} = - \cT_{\phi} \cL \cT_{\phi}^{-1} \quad ,\quad 
\cT_{\phi} \equiv \phi \cD_{+} \phi^{-1} \quad ,\quad \phi \equiv \P_{i_0}
\lab{DB-N2-def} \\
{\widehat \cL} = \bigl(\cT_{\psi}^{-1}\bigr)^\ast \cL \cT_{\psi}^\ast \quad ,\quad 
\cT_{\psi} \equiv \psi \cD_{+} \psi^{-1} \quad ,\quad \psi \equiv \Psi_{i_0}
\lab{adj-DB-N2-def}
\er
where $\P_{i_0}$ and $\Psi_{i_0}$ are some fixed bosonic (adjoint) SEF's entering
the pseudo-differential part of the $N\! =\! 2$ super-Lax operator
\rf{Lax-SKP-N2}. The (adjoint) DB-transformed super-Lax operator is of the
same form as the initial one:
\be
{\wti \cL} = \cD_{-} + \sum_{i=1}^M {\wti \P}_i \cD_{+}^{-1} {\wti \Psi}_i
\quad ,\quad
{\widehat \cL} = \cD_{-} + 
\sum_{i=1}^M {\widehat \P}_i \cD_{+}^{-1} {\widehat \Psi}_i
\lab{DB-Lax-SKP-N2}
\ee
\vspace{-.5cm}
\br
{\wti \P}_{i_0} = - \cT_{\phi}\(\cL (\P_{i_0})\) \;\; ,\;\;
{\wti \Psi}_{i_0} = {1\o {\P_{i_0}}}       \quad ;\quad
{\widehat \P}_{i_0} = - {1\o {\Psi_{i_0}}} \;\; ,\;\;
{\widehat \Psi}_{i_0} = \cT_{\psi}\(\cL^\ast (\Psi_{i_0})\)
\lab{DB-SEF-0}\\
{\wti \P}_i = - \cT_{\phi} (\P_i) \quad ,\quad
{\wti \Psi}_i = (-1)^{|i|}\bigl(\cT_{\phi}^{-1}\bigr)^\ast (\Psi_i) 
\quad {\rm for}\; i \neq i_0
\lab{DB-SEF}\\
{\widehat \P}_i = (-1)^{|i|}\bigl(\cT_{\psi}^{-1}\bigr)^\ast (\P_i)  \quad ,\quad
{\widehat \Psi}_i = \cT_{\psi} (\Psi_i) \quad {\rm for}\; i \neq i_0
\lab{adj-DB-SEF}
\er
(recall $\phi \equiv \P_{i_0},\, \psi \equiv \Psi_{i_0}$).
Note that the (adjoint) DB-transformed SEF's for all $i\neq i_0$ possess
opposite Grassmann parities w.r.t. the initial SEF's. Further we can show,
following the method of ref.\ct{svirflow}, that the super-DB transformations
\rf{DB-N2-def}--\rf{adj-DB-SEF} preserve (upto an overall sign change of the
fermionic flows) the subalgebra $\({\widehat {GL}}(M_B -1,M_F)\)_{+}$ of the
additional symmetry algebra $\({\widehat {GL}}(M_B,M_F)\)_{+}$
\rf{super-KM-alg-flows}.

Using relations \rf{N2-supertau} we deduce the following (adjoint) DB
transformations for the pertinent $N\! =\! 2$ super-tau function:
\be
\t \longrightarrow {\wti \t} = \frac{\phi}{\t} \quad, \quad
\t \longrightarrow {\widehat \t} = -\frac{1}{\psi\,\t}
\lab{DB-supertau}
\ee
Relations \rf{DB-supertau} are similar to those in the $N\! =\! 1$ super-KP case
\ct{match,svirflow} and have to be contrasted with their counterparts in the
ordinary ``bosonic'' case where
${\wti \t} = \phi\,\t\; ,\; {\widehat \t} = -\psi\,\t$.

Arbitrary iterations of any number of successive DB and adjoint-DB
transformations, {\sl i.e.}, general DB orbits of $N\! =\! 1$ constrained 
super-KP hierarchies have been worked out in detail in ref.\ct{svirflow}. 
The current $N\! =\! 2$ super-DB transformations 
\rf{DB-Lax-SKP-N2}--\rf{adj-DB-SEF} and \rf{DB-supertau}
have a structure analogous to their counterparts in the $N\! =\! 1$ super-KP 
case, with the following simple modifications: (i) the $N\! =\! 1$ 
super-derivative operator $\cD = \pa/\pa\th +\th \pa$ is replaced by the first
$N\! =\! 2$ super-derivative operator $\cD_{+}$; (ii) the $N\! =\! 1$ super-Lax
operator is replaced by the $N\! =\! 2$ super-Lax operator containing the second
$N\! =\! 2$ super-derivative operator $\cD_{-}$ in its positive differential 
part (cf. \rf{Lax-SKP-N2}). Therefore, we can easily generalize the results in 
\ct{svirflow} to obtain for the general DB-orbit of the $N\! =\! 2$ super-tau 
function the following Berezinian (super-determinant) expressions: 
\br
\frac{\t^{(0;0)}}{\t^{(n+2m;n)}} = (-1)^{mn + n(n-1)/2}
\times \phantom{aaaaaaaaaaaaaaaaaa}
\nonu \\
\phantom{aaaaaaaaaaaaaaaaaa}
\lab{DB-supertau-2m+n} \\
{\rm Ber} \threemat{
{\wti W}^{(n+m;m)}_{n+m,n+m} \llb \{\vp\} \bv \{\psi\}\rrb}{|}{
{\wti W}^{(m;n)}_{m,n+m} \llb \{\vp_{(\h)}\}\bv \{\psi\}\rrb
}{-------------}{|}{------------}{ 
W_{n+m,m} \lb \cD_{+}\vp_0 ,\ldots ,\cD_{+} \vp_{n+m-1}\rb}{|}{
W_{m,m} \lb \cD_{+}\vp_\h ,\ldots ,\cD_{+}\vp_{m-\h}\rb}
\nonu
\er
\vspace{.1in}
\br
\t^{(n+2m+1;n)}\, \t^{(0;0)} = (-1)^{mn + n(n-1)/2}  
\times \phantom{aaaaaaaaaaaaaaaaaa}
\nonu \\
\phantom{aaaaaaaaaaaaaaaaaa}
\lab{DB-supertau-2m+1+n} \\
{\rm Ber} \threemat{
{\wti W}^{(n+m+1;n)}_{n+m+1,n+m+1} \llb \{\vp\},\vp_{n+m}\bv \{\psi\}\rrb}{|}{
{\wti W}^{(m;n)}_{m,n+m+1} \llb \{\vp_{(\h)}\}\bv \{\psi\} \rrb
}{--------------}{|}{------------}{ 
W_{n+m+1,m} \llb \cD_{+}\vp_0,\ldots ,\cD_{+}\vp_{n+m} \rrb}{|}{
W_{m,m} \llb \cD_{+}\vp_\h ,\ldots ,\cD_{+}\vp_{m-\h}\rrb}
\nonu
\er

In Eqs.\rf{DB-supertau-2m+n}--\rf{DB-supertau-2m+1+n} the notations used are
as follows. The superscript $(k;l)$ of the super-tau function $\t^{(k;l)}$ 
indicates $k$ steps of DB transformations plus $l$ steps of adjoint-DB
transformations according to \rf{DB-Lax-SKP-N2}--\rf{adj-DB-SEF} and 
\rf{DB-supertau}. The block matrices entering the Berezinians in
\rf{DB-supertau-2m+n}--\rf{DB-supertau-2m+1+n} possess the following special
generalized Wronskian-like $k \times (m+n)$ matrix form:
\br
{\wti W}^{(k;n)}_{k,m+n} \llb \{\vp\} \bv \{\psi\} \rrb \equiv 
{\wti W}^{(k;n)}_{k,m+n} 
\llb \vp_0,\ldots ,\vp_{k-1} \bv \psi_{\h},\ldots ,\psi_{n-\h}\rrb =
\nonu \\
= \left(
\begin{array}{cccc}
\vp_0 & \cdots & \cdots & \vp_{k-1} \\
\vdots & \ddots & \ddots & \vdots   \\
\pa^{m-1}\vp_0 & \cdots & \cdots & \pa^{m-1}\vp_{k-1} \\
\cD_{+}^{-1}\! (\vp_0\psi_\h)& \cdots &\cdots &\cD_{+}^{-1}\! (\vp_{k-1}\psi_\h) \\
\vdots & \ddots & \ddots & \vdots   \\
\cD_{+}^{-1}\! (\vp_0\psi_{n-\h})& \cdots &\cdots
&\cD_{+}^{-1}\! (\vp_{k-1}\psi_{n-\h})    
\end{array} \right)    
\lab{susy-wti-Wronski}
\er
where $\{\vp\} \equiv \{\vp_0,\ldots ,\vp_{k-1}\}$ is a set of $k$ bosonic or
fermionic superfunctions whereas
$\{\psi\} \equiv \bigl\{\psi_\h,\ldots ,\psi_{n-\h}\bigr\}$ is a set of $n$
fermionic superfunctions. 
The generalized Wronskian-like matrix \rf{susy-wti-Wronski} is the $N\! =\! 2$
supersymmetric generalization of the Wronskian-like block matrices entering the 
general \DB determinant solutions for the tau-functions of ordinary ``bosonic''
constrained KP hierarchies \ct{hallifax,gauge-wz}. In the special case of $n=0$
\rf{susy-wti-Wronski} reduces to the rectangular $k \times m$ Wronskian matrix:
\be
W_{k,m} \llb \vp_0,\ldots ,\vp_{k-1} \rrb =
\left(
\begin{array}{cccc}
\vp_0 & \cdots & \cdots & \vp_{k-1} \\
\pa \vp_0 & \cdots & \cdots & \pa \vp_{k-1} \\
\vdots & \ddots & \ddots & \vdots   \\
\pa^{m-1}\vp_0 & \cdots & \cdots & \pa^{m-1}\vp_{k-1}
\end{array} \right)    
\lab{Wronski-matrix}
\ee
Accordingly, the sets of superfunctions entering the Wronskian-like matrix
blocks in the Berezinians \rf{DB-supertau-2m+n}--\rf{DB-supertau-2m+1+n}
are bosonic SEF's ($\{\vp\}\equiv \{\vp_0,\ldots ,\vp_{n+m-1}\}$),
fermionic SEF's ($\{\vp_{(\h)}\} \equiv \{\vp_\h,\ldots ,\vp_{m-\h}\} $)
and fermionic adjoint SEF's ($\{\psi\} \equiv \{\psi_\h,\ldots ,$
$\psi_{n-\h}\}$), 
respectively, which are expressed in terms of the constituent (adjoint) SEF's 
$\lcurl \P_i,\,\Psi_i\rcurl_{i=1}^M$ entering the pseudo-differential part of 
the super-Lax operator \rf{Lax-SKP-N2} in the following way:
\br
\{\vp\} = \(\,\Bigl\{\cL^{2p_k}(\P_k)\Bigr\}_{k=1,\ldots ,M_B}^{p_k =0,1,\ldots}
\; ,\; \Bigl\{\cL^{2q_l +1}(\P_{M_B +l})\Bigr\}_{l=1,\ldots ,M_F}^{q_l =0,1,\ldots}
\,\)
\lab{SEF-boson-set}\\
\{\vp_{(\h)}\} = 
\(\,\Bigl\{\cL^{2r_k}(\P_{M_B +k})\Bigr\}_{k=1,\ldots ,M_F}^{r_k =0,1,\ldots}
\; ,\; \Bigl\{\cL^{2s_l +1}(\P_l)\Bigr\}_{l=1,\ldots ,M_B}^{s_l =0,1,\ldots}
\,\)
\lab{SEF-fermi-set}\\
\{\psi\} = 
\(\,\Bigl\{\bigl(\cL^{2r_k}\bigr)^\ast (\Psi_{M_B +k})
\Bigr\}_{k=1,\ldots ,M_F}^{r_k =0,1,\ldots}  \; ,\; 
\Bigl\{\bigl(\cL^{2s_l +1}\bigr)^\ast (\Psi_l)
\Bigr\}_{l=1,\ldots ,M_B}^{s_l =0,1,\ldots} \,\)
\lab{adj-SEF-fermi-set}
\er
(recall that $\lcurl \P_i,\,\Psi_i\rcurl_{i=1}^{M_B}$ are bosonic, whereas
$\lcurl \P_{M_B +i},\,\Psi_{M_B +i}\rcurl_{i=1}^{M_F}$ are fermionic).

In conclusion let us write down explicitly the general DB-solution for the
$N\! =\! 2$ super-tau function \rf{DB-supertau-2m+n}--\rf{DB-supertau-2m+1+n}
of the simplest member of constrained $N\! =\! 2$ super-KP hierarchies' class
\rf{Lax-SKP-N2} with $M=1$, {\sl i.e.},
$\cL = \cD_{-} + \P \cD_{+}^{-1}\Psi$ with ``free'' initial super-Lax operator
$\cL^{(0;0)} = \cD_{-}$. It turns out that the latter solutions coincide
with the DB-solutions previously obtained in refs.\ct{N2-SKP,Lecht-Sorin}.

Indeed, in this case one can show, taking into account
\rf{SEF-boson-set}--\rf{SEF-fermi-set}, that the Berezinian 
expressions \rf{DB-supertau-2m+n}--\rf{DB-supertau-2m+1+n} reduce to
\foot{In this simplest case of $N\! =\! 2$ super-KP hierarchy, similarly to
its bosonic and $N\! =\! 1$ super-KP counterparts, adjoint-DB
transformations correspond simply to a backward shift on th DB orbit.}:
\be
\bigl(\t^{(2m;0)}\bigr)^{-1} =  \phantom{aaaaaaaaaaaaaaaaaaaaaaaaa}
\lab{DB-supertau-2m}
\ee
\vspace{-.5cm}
\br
{\rm Ber} \threemat{
W_{m,m} \llb \P_0,\ldots ,\pa^{m-1}\P_0 \rrb}{|}{
W_{m,m} \llb \cD_{-}\P_0, \ldots ,\cD_{-}\pa^{m-1}\P_0 \rrb
}{-------------}{|}{---------------}{ 
W_{m,m} \llb \cD_{+}\P_0, \ldots ,\cD_{+}\pa^{m-1}\P_0 \rrb}{|}{
W_{m,m} \llb \cD_{+}\cD_{-}\P_0, \ldots ,\cD_{+}\cD_{-}\pa^{m-1}\P_0\rrb}
\nonu
\er
\be
\t^{(2m+1;0)} =  \phantom{aaaaaaaaaaaaaaaaaaaaaaaa}
\lab{DB-supertau-2m+1}
\ee
\vspace{-.5cm}
\br
{\rm Ber} \threemat{  
W_{m+1,m+1} \llb \P_0,\ldots ,\pa^{m}\P_0 \rrb}{|}{
W_{m,m+1} \llb \cD_{-}\P_0, \ldots ,\cD_{-}\pa^{m-1}\P_0 \rrb
}{-------------}{|}{---------------}{ 
W_{m+1,m} \llb \cD_{+}\P_0, \ldots ,\cD_{+}\pa^{m}\P_0 \rrb}{|}{
W_{m,m} \llb \cD_{+}\cD_{-}\P_0, \ldots ,\cD_{+}\cD_{-}\pa^{m-1}\P_0\rrb}
\nonu
\er
where notation \rf{Wronski-matrix} for the pertinent Wronskian matrix blocks
has been used. In Eqs.\rf{DB-supertau-2m}--\rf{DB-supertau-2m+1} $\P_0$
denotes the ``free'' SEF (cf. \rf{super-spec} and \rf{xi-def}) :
\br
\P_0 (t, \th_{\pm}, \rho^{\pm}) = \int\! d\l\, d\eta_{\pm}
\phi_0 (\l,\eta_{\pm}) e^{\xi (t, \th_{\pm}, \pm\rho^{\pm};\l,\eta_{\pm})}
\lab{free-SEF} \\
\phi_0 (\l,\eta_{\pm}) = \phi^{(1)}_B (\l) + \eta_{+} \phi^{(1)}_F (\l)
+ \eta_{-} \bigl( \phi^{(2)}_F (\l) + \eta_{+} \phi^{(2)}_B (\l)\bigr)
\lab{free-spec}
\er
with an arbitrary $N\! =\! 2$ superspace ``density'' \rf{free-spec}. As in
the bosonic and $N\! =\! 1$ cases, supersymmetric ``soliton''-like solutions
for the $N\! =\! 2$ super-tau function 
\rf{DB-supertau-2m}--\rf{DB-supertau-2m+1} are obtained by choosing
delta-function form for the components of the $N\! =\! 2$ superspace ``density''
\rf{free-spec} :
\be
\phi^{(1,2)}_B = \sum_{a=1}^{N_B} c^{(1,2)}_a \d (\l - \l_a) \quad ,\quad
\phi^{(1,2)}_F = \sum_{a=1}^{N_F} \vareps^{(1,2)}_a \d (\l - \l_a)
\lab{free-spec-sol}
\ee
with arbitrary bosonic $\bigl\{ c^{(1,2)}_a,\l_a\bigr\}$ and fermionic
$\bigl\{\vareps^{(1,2)}_a\bigr\}$ constant parameters, and $N_{B,F}$ being
arbitrary integers.

\nonumsection{Outlook}
\noindent
The present results open a number of interesting problems for further
research, in particular: 

(i) Search for a negative-grade counterpart of the
positive-grade superloop superalgebra $\({\widehat {GL}}(M_B,M_F)\)_{+}$
\rf{super-KM-alg-flows} of additional symmetries of $N\! =\! 2$
(constrained) super-KP hierarchies \rf{Lax-SKP-N2}, as it is the case for
bosonic and $N\! =\! 1$ super-KP hierarchies \ct{virflow,gauge-wz,svirflow}.

(ii) Search for Virasoro (conformal) and super-Virasoro additional symmetries
of $N\! =\! 2$ super-KP hierarchies \rf{Lax-SKP-N2} (for the derivation
of Virasoro additional symmetries of constrained bosonic and $N\! =\! 1$ 
super-KP hierarchies, see refs.\ct{virflow,svirflow}).

(iiI) Detailed study of {\em extended} $N\! =\! 2$ super-KP hierarchies, obtained 
from the \cSKPe hierarchies \rf{Lax-SKP-N2} by adding to the set of original 
isospectral flows $\partder{}{t_l},\, D^{\pm}_n$, additional Manin-Radul-type 
subsets of (anti-)commuting additional symmetry flows, as it has been done in the 
$N\! =\! 1$ super-KP case \ct{svirflow}. Similarly to the bosonic and $N\! =\! 1$
super-KP cases (see refs.\ct{hallifax,gauge-wz,svirflow})
we can view the above extended $N\! =\! 2$ super-KP hierarchies as 
{\em matrix} generalizations of the original scalar \cSKPe hierarchies
\rf{Lax-SKP-N2}. Furthermore, as in the $N\! =\! 1$ super-KP case we can look 
for $N\! =\! 2$ supersymmetric generalizations of various interesting 
higher-dimensional nonlinear evolution equations (such as Davey-Stewartson 
nonlinear system) which are contained in the extended (matrix) $N\! =\! 2$ 
super-KP hierarchies.

(iv) Detailed study of the properties and possible physical significance of
the very broad class of \DB solutions 
\rf{DB-supertau-2m+n}--\rf{DB-supertau-2m+1+n}, in particular 
\rf{DB-supertau-2m}--\rf{DB-supertau-2m+1}.


\nonumsection{Acknowledgements}
\noindent
We gratefully acknowledge support from U.S. National Science Foundation 
grant {\sl INT-9724747}. We thank Prof. H. Aratyn and the Physics Department
of University of Illinois at Chicago for hospitality during the initial
stage of the project. This work is also partially supported by Bulgarian NSF 
grant {\sl F-904/99}. 

\nonumsection{References}

\end{document}